\documentclass[final]{aipproc}
  
\layoutstyle{8x11single}
\usepackage{aas_macros}
\def \etal{{\em et\ al.},\ }

\begin{document}

\title{Catching GRBs with atmospheric Cherenkov telescopes}

                \texttt{http://www.aip..org/pacs/index.html}

\author{R. C. Gilmore}{
  address={SISSA, via Bonomea 265, 34136 Trieste, Italy}
}

\author{J. R. Primack}{
  address={University of California, Santa Cruz, CA 95064 USA}
}

\author{A. Bouvier}{
  address={University of California, Santa Cruz, CA 95064 USA}
}

\author{A. N. Otte}{
  address={University of California, Santa Cruz, CA 95064 USA}
}


\begin{abstract}
Fermi has shown GRBs to be a source of $>$10 GeV photons.  We present an estimate of the detection rate of GRBs with a next generation Cherenkov telescope.  Our predictions are based on the observed properties of GRBs detected by Fermi, combined with the spectral properties and redshift determinations for the bursts population by instruments operating at lower energies.  While detection of VHE emission from GRBs has eluded ground-based instruments thus far, our results suggest that ground-based detection may be within reach of the proposed Cherenkov Telescope Array (CTA), albeit with a low rate, $0.25 - 0.5$/yr.  Such a detection would help constrain the emission mechanism of gamma-ray emission from GRBs.  Photons at these energies from distant GRBs are affected by the UV-optical background light, and a ground-based detection could also provide a valuable probe of the Extragalactic Background Light (EBL) in place at high redshift.

\end{abstract}

\maketitle


\section{Introduction}
The observation of gamma-ray bursts (GRBs) with ground-based imaging atmospheric Cherenkov telescopes (IACTs) has been a tantalizing possibility in recent years.  Powerful $>$10-meter telescope arrays such as H.E.S.S., MAGIC, and VERITAS have come online in the last decade, and satellite detectors such as Swift are capable of providing the necessary localization of events within seconds.  Despite major campaigns to respond to satellite burst alerts at all three of these instruments (e.g., \cite{aharonian09,albert07d,horan08}), no conclusive detection of a GRB with an IACT has yet been made.  However, observations of GRBs with the LAT instrument on the Fermi satellite have provided new insight into the emission of GRBs in the VHE band.  Since its launch on June 11, 2008, Fermi LAT has detected GRBs at $>$100 MeV energies at a rate of about 10/yr.  The highest energy detected photons have had rest frame energies exceeding 90 GeV.

In a recent paper \cite{gilmoreGRB} (GPP10), we addressed the possibility of detecting GeV-scale emission from GRBs with the Fermi LAT and MAGIC telescopes.  Using simple assumptions about the GRB rate and spectral extrapolation based on the Swift GRB population, this work predicted the number of GRBs that could be detected per year, and the likely number of photons events per background for each case.  GPP10 predicted the Fermi LAT to have a detection rate of 3-4 GRBs/yr above 10 GeV per year, a rate that was matched well in the first year of telescope operations, but may have since proven to be an overestimate.  The detection rate for MAGIC, ignoring instrumental background, was calculated to be 0.2-0.3 events per year, after somewhat optimistic assumptions were made about the telescope response time to satellite burst triggers.  While the rate predicted for MAGIC was low, and may explain the lack of detections by the current generation of ground-based instruments, the predicted gamma-ray rates for a GRB detection near zenith were found to be up to thousands of counts in the lower energy range of the experiment within the prompt and early afterglow phases of the GRB, a large potential scientific payoff.

Here, we describe a new calculation with a more sophisticated treatment of the spectral extrapolation to high energies and the telescope detection capabilities.  We have also implemented the $t^{-1.5}$ afterglow feature that has been seen in several GRBs by LAT \cite{ghisellini10}, and may more realistically represent the signal visible at GeV energies.  In the next section, we describe the ingredients and motivation of our model for high energy emission.  

\section{Methods}
\label{sec:methods}
To a large extent, the challenge of modeling GRBs arises out of the large variance in properties seen between bursts, and the lack of a simple model describing the radiative mechanism.  In GPP10, our extrapolation of GRB emission to high energies was based on the assumption that flux at VHE energies could be described as a fixed fraction of the flux at lower energies.  In that work, a flux ratio of  \[ \frac{F(100 \mbox{ MeV} - 5 \mbox{ GeV})}{F(20\mbox{ keV} - \mbox{ 2 MeV})} = 0.1\] was assumed, reflecting the typical ratio inferred from coincident BATSE/EGRET observations.  
While photon statistics for EGRET-observed GRBs were severely limited, LAT observations have now given us an opportunity to reexamine our assumptions about these flux correlations. 

\begin{figure}
  \includegraphics[height=.35\textwidth]{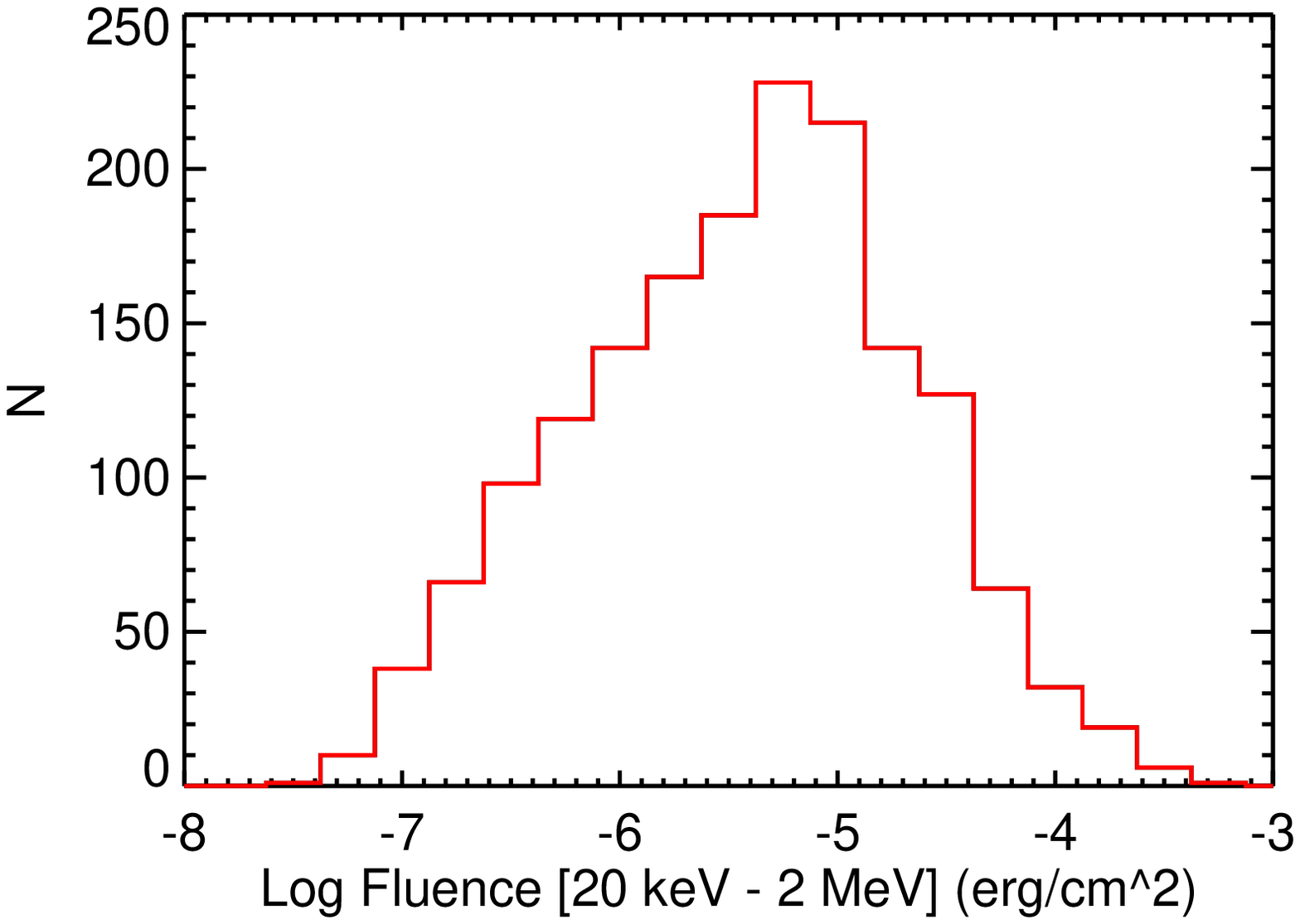}
  \includegraphics[height=.35\textwidth]{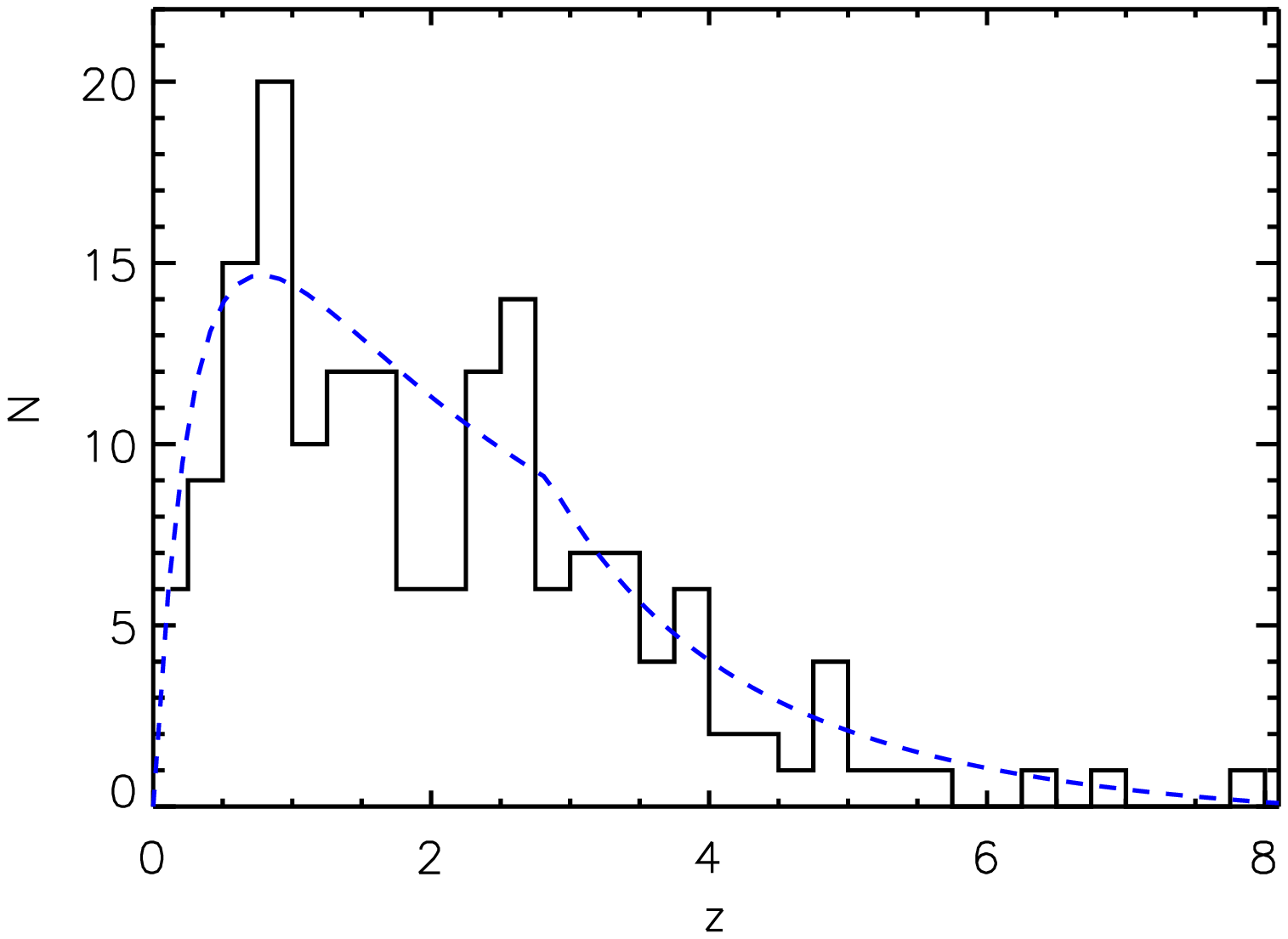}
\caption{{\bf Left:} Histogram of total fluence in the BATSE energy range for our sample of GRBs.  {\bf Right:} The redshifts determined for Swift GRBs.  The dashed line shows the fit used in this work.}
\label{fig:batsefluhist}
\end{figure} 
 
 In this work, we have turned to the large sample of GRBs available from the BATSE catalog, providing the GRB fluence distribution and Band \cite{band93} spectral fits used in our analysis which we briefly describe in this section.  The redshift distribution of GRBs is assumed to follow that of the Swift GRB population.  Figure 1 shows input distributions for fluence and redshift used in our work.   Attempting to build a sample of GRBs for analysis from populations of two different instruments will necessarily introduce biases.  These biases will be studied in a upcoming publication and included amongst a comparison of different populations.  Our calculation includes the impact of attenuation by UV-optical background photons on GRB emission using the fiducial model of \cite{gilmoreUV}.
 
 \subsection{Spectral extrapolation}
Each of the 4 bright GRBs seen by LAT above 10 GeV shows differing behavior.  GRB 080916C, the first such detection by LAT (seen about 3 months after launch), was found to be well-described in all time bins by a continuation of the Band function determined at GBM energies \cite{greiner09}.  Separate spectral components from the Band function were found to be required to match the GeV-scale emission of the 3 other brightest GRBs in the LAT catalog, long-duration GRBs 090902B and 090926, and short burst 090510.  We have therefore considered two methods of extrapolating the lower energy emission to higher energies in this work.  

As a minimal model, we consider the visibility predicted for GRBs at high energy without any significant deviation from the Band fit.   In this model, labeled `bandex' below, the high-energy spectrum is simply a continuation of the Band spectra determined at lower energy.  The high energy normalization is therefore determined by the Band function peak energy, normalization, and the upper energy index $\beta$, which continues to GeV energies.  We have enforced the requirement that $\beta$ not be harder than -2, and the bursts in the sample with harder spectra are reset to this value.


\begin{figure}
 \includegraphics[height=.2\textheight]{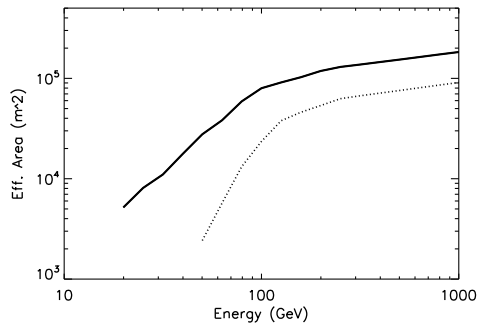}
 \caption{The effective area functions used in this work.  The dotted line is the effective area for the MAGIC telescope after all standard analysis cuts have been performed \cite{albert08a}.  The solid curve is the estimated effective area for the central cluster of telescopes in CTA used in this analysis.}
 \label{fig:ea}
\end{figure} 
 
In the fixed-parameter (``fixed'') model, we make the assumption that the fluence between BATSE (20 keV to 2 MeV) energies and GeV-scale (100 MeV to 10 GeV) energies can be described by a single ratio.  We use here a flux ratio of 0.1, which is well supported by observations of GBM-LAT observations of long-duration GRBs \cite{dermer10}.  The spectral index at high energies is set to -2, consistent with the mean value for EGRET GRBs of -1.95 \cite{dingus95,le&dermer09}, and near the center of the distribution for LAT-detected events \cite{ghisellini10}.  Both of these values are quite similar to the prompt emission assumptions used in GPP10.   In general, this model requires a significant departure from the extrapolated Band function, and implies the appearance of a separate high-energy spectral component.

\subsection{Telescope Properties}
As many of the properties of the CTA are indeterminate at the time of writing, we have relied on the design concept for the array described in \cite{ctaconcept10}, as well as reasonable extrapolations from the current generation of IACTs, particularly the MAGIC telescope.  Our assumption about the effective area function of CTA, shown here in Figure \ref{fig:ea}, is based on configuration E, which assumes a central cluster of four 24-meter class telescopes that determine the low-energy sensitivity of the instrument.  Sensitivity at energies above a few hundred GeV, which is provided by more dispersed arrays of 12- and 7-meter class instruments, is not crucial to our results here, as most GRBs will occur at redshifts for which emission at these energies is strong attenuated by the EBL.  

The transient and random nature of GRB emission represent the main difficulty in detecting emission from these sources.  The satellite localization time of the event, transmission of the data to the ground, and slew time for the IACT all contribute to a total delay time for the commencement of observation.    The localization time is dependent on the instrument and brightness of the GRB, but times of $<$ 15 sec are typical, and the transmission time is expected to be nearly instantaneous \cite{bastieri05}.  The largest class of telescopes in CTA, which provide coverage at the crucial low energies, are expected to have a slew time of 20 to 30 seconds.  As a baseline assumption, we assume a total response time of 60 seconds in this work.

\section{Discussion}

The GRB detection capabilities of the CTA can be described as the product of two independent factors: the detection efficiency, or probability that a randomly-selected GRB from our population occurring within the field-of-view of the telescope produces a detectable signal, and the rate at which the telescope is able to respond to triggers from satellite instruments within that field of view.  Our modeling of the former is done by simulating observations from the population of GRBs described in the previous section.  For each GRB, we determine the integrated photon counts and expected background counts on timescales varying from 1 to $10^4$ sec.  The significance of the detection is then calculated using the procedure described in  \cite{li&ma83}.  The GRB is then assumed to be detected if the significance is higher than 5$\sigma$ and at least 10 photons were seen above background in at least one timescale bin.

\begin{figure}
  \includegraphics[height=.33\textwidth]{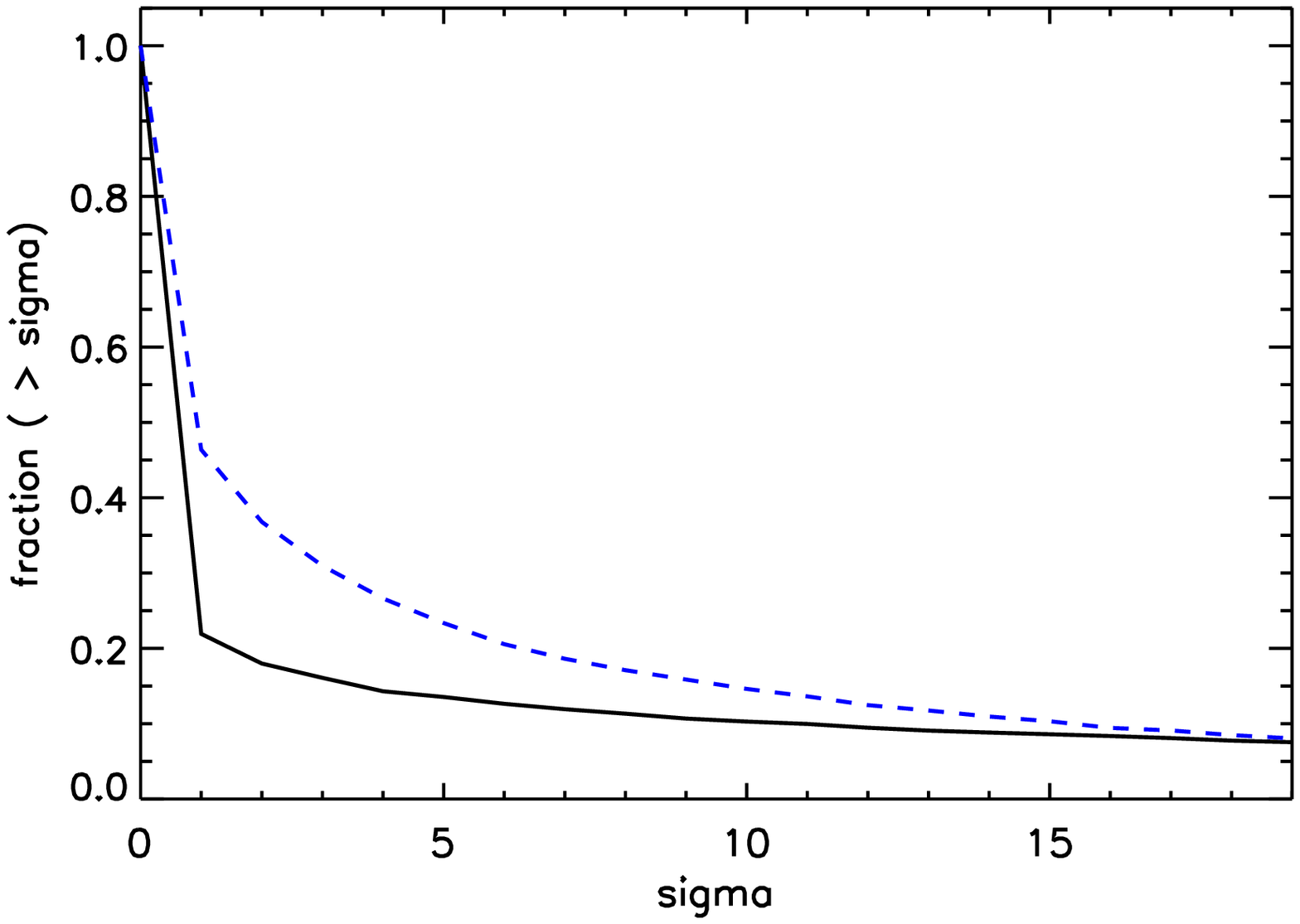}  
  \includegraphics[height=.33\textwidth]{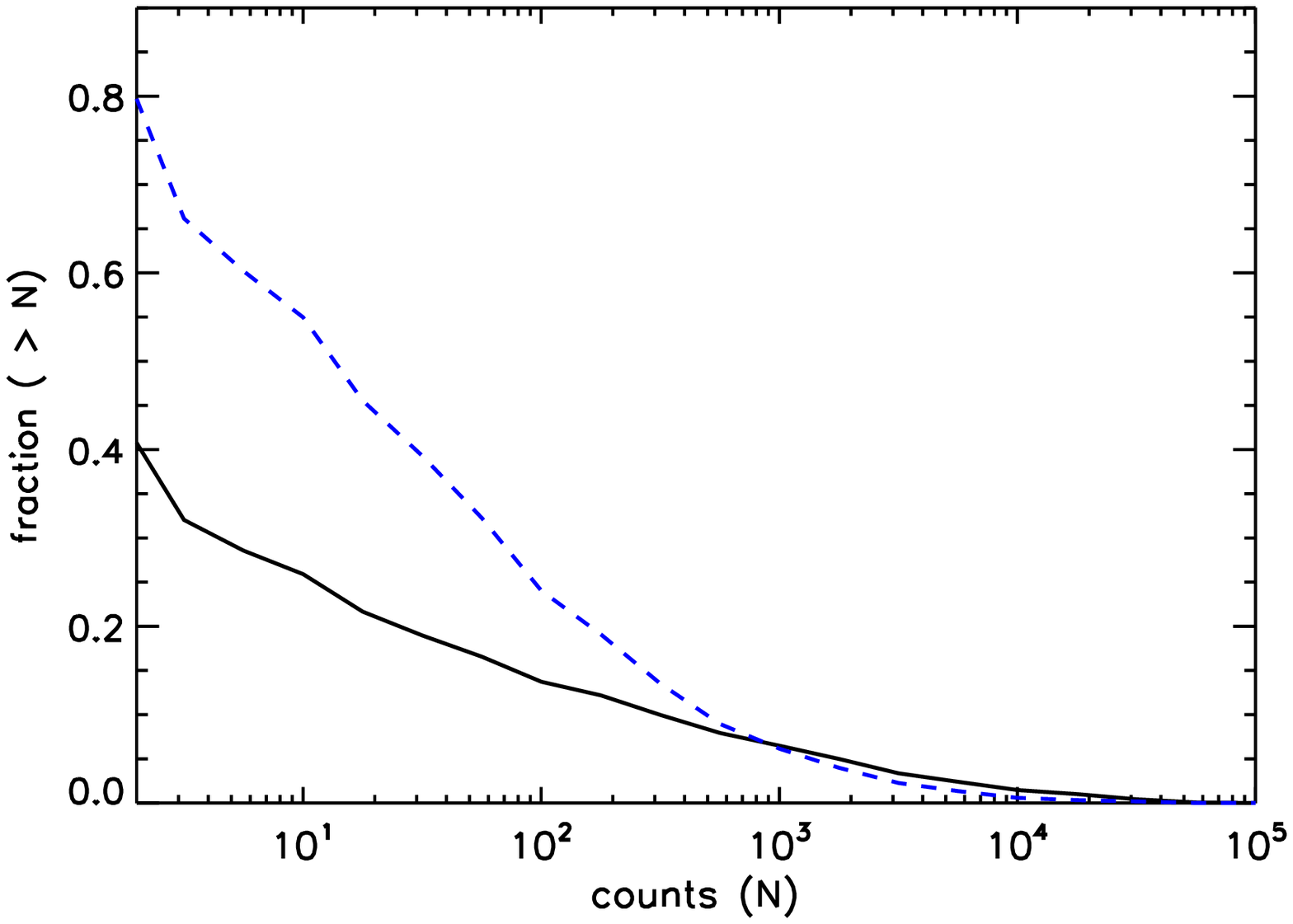}
\caption{{\bf Left:} The integral distribution of sigma values (significance of detection) for GRBs in our population.  The solid line is for the direct extrapolation of Band functions (`bandex' model), from the distribution seen in BATSE GRBs, and the broken line is for the `fixed model', using parameters described in the last section.  {\bf Right:} The integral distribution of counts above background, in the timescale bin with maximum sigma for each GRB.  Note that the y-axis intercept on this scale is the probability of $\geq 1$ count.}
\label{fig:ctasigma}
\end{figure}

In Figure \ref{fig:ctasigma}, we show results for an analysis done for a field of view $\theta_{\mbox{\footnotesize zenith}} < 45$ deg, i.e., assuming the GRB occurs randomly on the sky within 45 deg of zenith.  The curves in each plot are the integral probability distribution for the significance of the recorded gamma-ray signal, and the number of photons above background.


This baseline model predicts a detection efficiency of GRBs of 14\% and 23\% in the `bandex' and `fixed' models, respectively.  While these two models do predict somewhat different distributions in terms of significances for recorded signal and absolute photon counts, the predictions for detection efficiency using our above criterion (N$_\gamma \geq 10 \, ; \, \sigma \geq 5$) differ by less than a factor of 2.

An approximate prediction for the trigger rate can be easily calculated using the alert rate from satellite instruments.  The Swift satellite has historically  observed GRBs at a rate of about 95/yr \cite{gehrels09}.  With a 10\% duty cycle and a correction for the anti-solar bias present in Swift alerts (see discussion in \cite{gilmoreGRB}), a rate of $\sim$2 GRBs per year within 45 degrees of zenith can be estimated.  Our results suggest that a diligent GRB observation campaign with the CTA could detect GRBs at a rate of one every 2--4 years, with great scientific impact.  However we note that the presence of a GeV-scale spectral cutoff, which we have not considered here, could reduce this rate significantly.

In this proceeding we have described a method for determining the detection probability of GRBs with ground-based IACT instruments.  An upcoming publication will include a detailed presentation of our model and results for the CTA experiment, including calculating the hypothetical effect of varying different telescope parameters, such as response time and low-energy effective area.  Alerts from other satellite instruments, including Fermi GBM and the upcoming SVOM experiment, could contribute to the trigger rate and will be considered as well.


\begin{theacknowledgments}
RCG acknowledges a research fellowship from the SISSA Astrophysics Sector.  RCG and JRP acknowledge Fermi Gamma Ray Telescope Theory grants.
\end{theacknowledgments}



\bibliographystyle{aipproc}   



\end{document}